\documentclass[prl,twocolumn,amsmath,amssymb]{revtex4-1}
\usepackage{dcolumn}% Align table columns on decimal point
\usepackage{bm}% bold math
\usepackage{epsfig}
\usepackage{graphicx}% Include figure files
\usepackage{amsmath}
\usepackage{amssymb}
\usepackage{mathrsfs}
\usepackage{color}
\usepackage[usenames,dvipsnames]{xcolor}
\usepackage{hyperref}
\usepackage[caption=false]{subfig}
\usepackage{amsfonts}
\usepackage{braket}

\def\be{\begin{equation}}
\def\ee{\end{equation}}
\def\bea{\begin{eqnarray}}
\def\eea{\end{eqnarray}}
\def\nn{\nonumber}

\def\slashchar#1{\setbox0=\hbox{$#1$}           % set a box for #1
   \dimen0=\wd0                                 % and get its size
   \setbox1=\hbox{/} \dimen1=\wd1               % get size of /
   \ifdim\dimen0>\dimen1                        % #1 is bigger
      \rlap{\hbox to \dimen0{\hfil/\hfil}}      % so center / in box
      #1                                        % and print #1
   \else                                        % / is bigger
      \rlap{\hbox to \dimen1{\hfil$#1$\hfil}}   % so center #1
      /                                         % and print /
   \fi}

\begin{document}

\title{Symmetry from Entanglement Suppression}

%\vspace*{-0.2cm}

\author{
%\vspace{0.5cm}
\mbox{Ian Low$^{\,a,b}$ and Thomas Mehen$^{\,c}$}
 }
\affiliation{
%\vspace*{.5cm}
$^a$\mbox{{High Energy Physics Division, Argonne National Laboratory, Argonne, IL 60439, USA}}\\
$^b$\mbox{{Department of Physics and Astronomy, Northwestern University, Evanston, IL 60208, USA}} \\
$^c$\mbox{{Department of Physics, Duke University, Durham, NC 27708, USA}}%\\
}

\begin{abstract}
Symmetry is among the most fundamental and powerful concepts in nature, whose existence is usually taken as given, without explanation. We explore whether symmetry can be derived from more fundamental principles from the perspective of quantum information. Starting with a two-qubit system, we show  there are only two minimally entangling logic gates:  the Identity and the ${\text{\small SWAP}}$, where ${\text{\small SWAP}}$ interchanges the states of the qubits.  We further demonstrate that, when viewed as an entanglement operator in the spin-space, the $S$-matrix in the two-body scattering of fermions in the $s$-wave channel is uniquely determined by unitarity and rotational invariance to be a linear combination of the Identity and the ${\text{\small SWAP}}$.  Realizing a minimally entangling $S$-matrix would give rise to global symmetries, as exemplified in  Wigner's spin-flavor symmetry and Schr\"odinger's conformal invariance in low energy Quantum Chromodynamics. For $N_q$ species of qubit, the Identity gate is associated with an $[SU(2)]^{N_q}$ symmetry, which is enlarged to $SU(2N_q)$ when there is a species-universal coupling constant.
\end{abstract}

\maketitle

\section{Introduction} 
 
Symmetry plays an essential role with respect to the laws of nature.   Mathematically defined as invariance under a specified group of transformations, the notion of symmetry has  provided unity to vastly different physical phenomena. For example, different physical systems with the same underlying symmetries have the same critical exponents at second order phase transitions. The existence of conserved quantities, such as  energy and momentum, is typically a consequence of an underlying symmetry, in this case  invariance under time and spatial translations. Effective field theories, including the Standard Model, are defined by writing the most general Lagrangian consistent with known  symmetries.  

But what is the origin of symmetry? In a physical theory symmetries are usually given as inputs at short distances and are valid over a large range of length scales. Sometimes symmetries could be accidental and emerge only at long distances, which nonetheless cannot be predicted a priori. It is not known whether symmetry can be the outgrowth of more fundamental principles.

On the other hand, in attempting to answer the question of existence, J. A. Wheeler adopted the philosophy that every physical quantity is derived from binary bits, inventing  the phrase {\em It from bit} \cite{Wheeler1989-WHEIPQ}.  Indeed, remarkable connections between fundamental physics and information science have been revealed in the past decades. Central to this pursuit is quantum entanglement, a basic phenomenon in quantum mechanics. Naturally one is led to wonder: can symmetry come from qubit? \footnote{In this work we will focus on global symmetries realized in the Wigner-Weyl mode, where the energy spectra of the system furnish linear representations of the underlying symmetry.}

In this regard  there is a fascinating observation in  low energy Quantum Chromodynamics (QCD) recently. Ref.~\cite{Beane:2018oxh} computed the entanglement power in two-nucleon scattering and found  entanglement suppression in regions of parameter space with enhanced Wigner's spin-flavor symmetry and Schr\"odinger's  non-relativistic conformal invariance. This  raises the intriguing possibility of understanding symmetry from quantum entanglement.

In this work we aim to purify the observation and study the correlation between symmetry and entanglement suppression in an information-theoretic setting. As a result, our conclusions are universal, quantum mechanically speaking, and not confined to QCD. 

\section{Alice and Bob Meet Cartan}

We start by considering two qubits, Alice $(A)$ and Bob ($B$), each with the basis vectors $|0\rangle$ and $|1\rangle$. For a spin-1/2 system, $|0\rangle$ and $|1\rangle$ can be identified with states having the spin-component $S_z=+{1}/{2}$ and $-{1}/{2}$, respectively.
It is customary to define the computational basis $\{|00\rangle,|01\rangle,|10\rangle,|11\rangle\}$, where $|ij\rangle = |i\rangle_A\otimes |j\rangle_B$.  There are several measures of entanglement of two-qubit systems \cite{Kraus_2001} but our analysis relies only on general features of these measures, namely that they vanish if $|\Psi \rangle$ is unentangled, i.e., can be written as a direct product state, $|\Psi\rangle = |\phi\rangle\otimes |\phi^\prime \rangle$ for some $|\phi\rangle, |\phi^\prime \rangle$,  and are maximal for maximally entangled states. Examples of maximally entangled states include the Bell basis:
\be
|\Phi^\pm\rangle = \frac{\left(|00\rangle \pm |11\rangle\right)}{\sqrt{2}} \ , \quad  |\Psi^\pm\rangle = \frac{\left(|01\rangle \pm |10\rangle\right)}{\sqrt{2}}\ ,
\ee
as well as the magic basis \cite{PhysRevLett.78.5022}, defined in the same way as the Bell basis except for some phases: $|\Phi_1\rangle = |\Phi^+\rangle, |\Phi_2\rangle=-i|\Phi^-\rangle, |\Phi_3\rangle = |\Psi^+\rangle, |\Phi_4\rangle=-i|\Psi^-\rangle$. An example of an entanglement measure is  the concurrence $\Delta(\Psi)$ \cite{PhysRevLett.78.5022,PhysRevLett.80.2245}, which for a normalized state $|\Psi\rangle= \alpha |00\rangle +\beta |10\rangle +\gamma|01\rangle + \delta |11\rangle$ is defined as
\be
\label{eq:concur}
  \Delta (\Psi)= | \langle \Psi | \sigma_y \otimes \sigma_y|\Psi^*\rangle |=2|\alpha \delta-\beta \gamma|\ .
\ee
The concurrence has the property that $0\le \Delta(\Psi) \le 1$, where $\Delta(\Psi)=0$ if $|\Psi\rangle$ is not entangled  and $\Delta(\Psi)=1$ if it is maximally entangled.  Concurrence is reviewed in the Appendix and its relation to some other entanglement measures is discussed. The crucial property that all these measures share is that the entanglement is unchanged by single qubit unitary operators.

A quantum-mechanical operator acting on the two-qubit system is represented by a matrix $U$ in the $SU(4)$ group, which is parameterized by 15 generators:
\be
\{\bm{\sigma}^a\mbox{\small $\otimes$}\bm{1}_2,\ \bm{1}_2\mbox{\small $\otimes$} \bm{\sigma}^b, \ \bm{\sigma}^a\mbox{\small $\otimes$}\bm{\sigma}^b;\ a, b=x,y,z\} \ ,
\ee
where $\bm{\sigma}^a$'s are the Pauli matrices and $\bm{1}_2$ is the $2\times 2$ identity matrix.
It turns out that not all $U$'s generate entanglement; operators that are local in the product space, $U=U_A\mbox{\small $\otimes$} U_B$, $U_{A/B} \in SU(2)$,  leave the entanglement unchanged, leading to the notion of equivalent classes: two operators $U$ and $U^\prime$ generate the same amount of entanglement if they are related by local operators,
\be
U \sim U^\prime , \quad {\rm if} \quad U = (U_A\mbox{\small $\otimes$} U_B) U^\prime (V_A\mbox{\small $\otimes$} V_B) \ ,
\ee
for some single qubit operators $U_{A/B}$ and $V_{A/B}$.

Classification of all non-local, and hence entanglement generating operators in a two-qubit system has been achieved long  ago \cite{2000quant.ph.10100K,Kraus_2001,Zhang_2003}  and relies on the Cartan decomposition \cite{helgason2001differential} of the Lie algebra of a symmetric coset space $G/H$, whose generators satisfy
\be
\label{eq:symmetric}
[\mathfrak{h}, \mathfrak{h}]=\mathfrak{h} \ ,\ \    [\mathfrak{h}, \mathfrak{p}]=\mathfrak{p}\ ,\ \   [\mathfrak{p}, \mathfrak{p}]=\mathfrak{h}\ ,
\ee
where $\mathfrak{h}=\{T^i\}$ and $\mathfrak{p}=\{X^a\}$ are generators of $H$  and $G/H$, respectively. The Lie algebra $\mathfrak{g}$ of $G$ can be written as $\mathfrak{g}=\mathfrak{h}\oplus \mathfrak{p}$. Let $\mathfrak{a}=\{A^a\}$  be the maximal abelian subalgebra of $\mathfrak{g}$ contained in $\mathfrak{p}$: $\mathfrak{a}\subset \mathfrak{p}$ and $[\mathfrak{a},\mathfrak{a}]=0$. Then an element $g$ of $G$ can be written as
\be
\label{eq:cartandecom}
g = K_1\, e^{i\beta_a A^a} \, K_2 \ , \qquad \ K_1, K_2 \subset H \ ,
\ee
for some real parameters $\{\beta_a\}$. This is generally referred to as the Cartan decomposition for symmetric spaces.

The Cartan decomposition in Eq.~(\ref{eq:cartandecom}) can be understood as follows. The coset structure allows for the ``polar decomposition" for any $g\in G$,
\be
\label{eq:polar}
g= \bar{K}_1 L \ , \qquad \bar{K}_1 \in H \ \  \text{and}\ \ L \in G/H \ .
\ee
Next we would like to show that,
\be
\label{eq:pha}
\mathfrak{p} = \mathfrak{a} \oplus [ \mathfrak{h}, \mathfrak{a}]\ .
\ee
Recall that, since $\mathfrak{a}\subset \mathfrak{p}$,  Eq.~(\ref{eq:symmetric}) implies $[\mathfrak{h}, \mathfrak{a}] \subset \mathfrak{p}$. Consider a generator $X^a\in \mathfrak{p}$ that is orthogonal to $[\mathfrak{h},\mathfrak{a}]$,
\be
\label{eq:spaneq}
0 ={\rm Tr}\left(X^a [\mathfrak{h}, \mathfrak{a}]\right)  = {\rm Tr}\left(\mathfrak{h} [ \mathfrak{a},X^a ]\right) \ ,
\ee
where we have used the cyclic property of the trace in the second equality. Using Eq.~(\ref{eq:symmetric}) again, $[ X^a, \mathfrak{a}]\subset \mathfrak{h}$ which implies the only way Eq.~(\ref{eq:spaneq}) can hold is if $[ X^a, \mathfrak{a}]=0$. This shows $X^a$ must be part of the maximal Cartan subalgebra $\mathfrak{a}\subset \mathfrak{p}$ and Eq.~(\ref{eq:pha}) follows. The last step is to apply Baker-Campbell-Hausdorff formula,
\be
\label{eq:hausdorf}
e^{iZ} \, e^{iY} \, e^{-iZ}=e^{i(Y+[Z,Y]+\frac1{2!}[Z,[Z,Y]]+\frac1{3!}[Z,[Z,[Z,Y]]]+\cdots)} \ ,
\ee
with $Y=\mathfrak{a}$ and $Z=\mathfrak{h}$. Using Eqs.~(\ref{eq:symmetric}) and (\ref{eq:pha}) we see
$[\mathfrak{h},[\mathfrak{h},\mathfrak{a}]]=\mathfrak{p}$, $[\mathfrak{h},[\mathfrak{h},[\mathfrak{h},\mathfrak{a}]]]=\mathfrak{p}$, etc. Therefore the right-hand side of Eq.~(\ref{eq:hausdorf}) sits in $G/H$, leading to
\be
\label{eq:peae}
L =( K_2)^{-1}\, e^{i\beta_a A^a} K_2\ .
\ee
Plugging Eq.~(\ref{eq:peae}) into Eq.~(\ref{eq:polar}) we arrive at  Eq.~(\ref{eq:cartandecom}).

\section{Minimal Entanglers}

For  $SU(4)/(SU(2)\otimes SU(2)$), with the $SU(2)$'s generated by $\bm{\sigma}^a\mbox{\small $\otimes$}\bm{1}_2$ and $\bm{1}_2\mbox{\small $\otimes$} \bm{\sigma}^b$,
 $\mathfrak{a}=\{\bm{\sigma}^a\mbox{\small $\otimes$} \bm{\sigma}^a,\  a=x,y,z\}$. The space of inequivalent ``entanglers" is then parameterized by only 3 parameters, $(\beta_x, \beta_y,\beta_z)$, 
\bea
\label{eq:masterA}
\!\!\!\!\!\!\!\! U_d \equiv e^{i \beta_a \bm{\sigma}^a\mbox{\small $\otimes$} \bm{\sigma}^a}&=& \left(c_x c_y c_z+i\, s_x s_y s_z\right)\bm{1} \nonumber\\
 && +\left[ \left(s_z s_y c_x+i\, c_z c_y s_x\right) \bm{\sigma}^x\mbox{\small $\otimes$} \bm{\sigma}^x\right. \nonumber\\
 &&+\left.\left(\mbox{cyclic perm. of $\{x,y,z\}$} \right)\right] \, ,
\eea
where $c_a =\cos \beta_a$ and $s_a=\sin\beta_a$ and $\bm{1}$ is the four-dimensional identity matrix. It is worth recalling some basic properties of $U_d$. First, the amount of entanglement created by $U_d$ is periodic in $\beta_i\to \beta_i + \pi/2$ \cite{Kraus_2001}. To see this, consider 
\be
e^{i(\beta_x+\delta)\bm{\sigma}^x\mbox{\small $\otimes$} \bm{\sigma}^x} = \left( \bm{1} \cos \delta + i \, \bm{\sigma}^x\mbox{\small $\otimes$} \bm{\sigma}^x \sin \delta\right)e^{i\beta_x \bm{\sigma}^x\mbox{\small $\otimes$} \bm{\sigma}^x} \ .
\ee
When $\delta=\pi/2$ the above operator is related to the operator with $\delta=0$ by an operator that is a product of local unitaries, $U_A\otimes U_B=i \, \bm{\sigma}^x \otimes \bm{\sigma}^x$, and therefore belongs to the same equivalent class. Moreover, the eigenvalues of $U_d$ (most easily computed in the Bell basis which is the eigenbasis of $U_d$) are $\lambda_i = e^{i \phi_i}$, where $\phi_i$ are 
\begin{align}
 \label{phases}
\phi_1&= \beta_x-\beta_y+\beta_z \ ,& \quad 
\phi_2 &= -\beta_x+\beta_y+\beta_z  \nn \\
\phi_3 &= \beta_x +\beta_y-\beta_z  \ ,& \quad
\phi_4 &= -\beta_x - \beta_y - \beta_z \ .
\end{align}
These are the eigenvalues of $|\Phi^+\rangle$, $|\Phi^-\rangle$, $|\Psi^+\rangle$, and  $|\Psi^-\rangle$, respectively. Then $(\beta_a,\beta_b)\to (-\beta_a, -\beta_b)$ or $(\beta_a,\beta_b)\to (\beta_b,\beta_a)$ generate all possible permutations of the eigenvalues. Permuting the eigenvalues of $U_d$ does not change the entanglement properties of $U_d$. The geometric structure underlying all $U_d$ is a 3-torus, $T^3=S^1\times S^1\times S^1$, and we can restrict to $\pi/2 > \beta_x \ge \beta_y \ge \beta_z \geq 0$, $\pi/2 \ge \beta_x+\beta_y$. If $\beta_z =0$ we have the additional constraint $\pi/4 \ge \beta_x$. This corresponds to one Weyl chamber of $G/H$ \cite{Zhang_2003,2005quant.ph..7171T}, where every point corresponds to an equivalent class. We will use $[U_d]$ to denote the equivalent class containing $U_d$. Related discussions in the context of the $S$-matrix in a quantum field theory can be found in Ref.~\cite{Beane:2020wjl}.

 To quantify the ability of an operator to produce entanglement, it is possible to define the entanglement power of an operator by averaging over the states it acts on \cite{PhysRevA.63.040304}. For our purpose, it is sufficient to consider  $U_d|\psi\rangle$, where $|\psi\rangle = (a|0\rangle +b |1\rangle)\otimes (c |0\rangle + d |1\rangle)$ is a general product state in the computational basis, compute the concurrence  defined in Eq.~(\ref{eq:concur}) and average over the wave function of the qubit on the Bloch sphere. It turns out that it is easier to compute the average of the square of the concurrence ~\cite{BallardWu2011,PhysRevA.70.052313}, as explained in the Supplementary Material,
 \be
 \label{eq:enpower}
\frac12 \overline{ \Delta(U_d|\psi\rangle)^2} =  \frac{1}{6} -\frac{1}{18}\sum_{a<b} \cos(4 \beta_a ) \cos(4 \beta_b ) \, .
\ee
Since Eq.~(\ref{eq:enpower}) is positive-definite, a vanishing entanglement power  requires $\cos(4 \beta_a ) \cos(4 \beta_b )=1$ for all $a$ and $b$. The only solutions are $\beta_x=\beta_y=\beta_z =0$ or $\pi/4$. So there are only two minimally entangling $U_d$ whose matrix representations  in the computational basis are, 
\bea
\label{eq:idengate}
\!\!\!\!\!\!\!\!\!\!\!\!  \beta_a =0&:& U_d= \bm{1} \ , \nn \\
\!\!\!\!\!\!\!\!\!\!\!\!  \beta_a =\frac{\pi}{4}&:& U_d= e^{i\frac{\pi}4} \left[\begin{array}{cccc} 1 & 0 & 0 & 0 \\
                                         0  & 0 & 1 & 0 \\
                                         0  & 1 & 0 & 0 \\
                                         0  & 0 & 0 & 1 \end{array}\right] \equiv e^{i\frac{\pi}4}\times(
                                         {\text{\small SWAP}})\, .
\eea
The ${\text{\small SWAP}}$ interchanges the states of qubit $A$ and qubit $B$, i.e.,  $|ij\rangle \to |ji\rangle$ in the computational basis and  ${\text{\small SWAP}}^2 = \bm{1}$. The corresponding equivalence classes are $[\bm{1}]$ and $[{\text{\small SWAP}}]$.

\section{Two-body Scattering} 

In a scattering experiment $AB\to AB$, where $A$ and $B$ are fermions, the $S$-matrix can be considered as an entanglement operator in the spin-space. Moreover, it is well-known that the low-energy scattering of non-identical fermions is dominated by the $s$-wave channel, which we  focus on here. See Refs.~\cite{HARSHMAN_2005,2006quant.ph..6011H} for prior discussions of entanglement in fermion scattering.

In the $s$-wave channel, there is no orbital angular momentum and the $S$-matrix, ${\cal S}$, can be decomposed into the singlet, $^1S_0$, and the triplet, $^3S_1$, channels \footnote{In the $^3S_1$ channel there is a small mixing with $^3D_1$ due to the spin-orbital coupling, which we ignore.}, ${\cal S}={\cal S}_0+{\cal S}_1$. The general form of ${\cal S}$ is constrained by 1) unitarity: ${\cal S}^\dagger {\cal S}=1$, and 2) rotational invariance: $[{\cal S}_i, \vec{J}\,]=0$, where in the $s$-wave $\vec{J}=\vec{s}_A+\vec{s}_B$ is the total spin of $A$ and $B$.  In the computational basis, $\vec{s}_A=(\vec{\sigma}/2)\otimes\bm{1}_2$ and $\vec{s}_B=\bm{1}_2\otimes(\vec{\sigma}/2)$. Therefore ${\cal S}_i$ can only be a linear combination of the identity $\bm{1}$ and $J^2=(\vec{s}_A+\vec{s}_B)^2$: 
\be
{\cal S}_i=e^{i2\delta_i} \left(a_i \, \bm{1} + b_i \, \bm{\sigma}\cdot \bm{\sigma} \right) \ , \quad \bm{\sigma}\cdot \bm{\sigma} \equiv \sum_{a} \bm{\sigma}^a\otimes\bm{\sigma}^a \ .
\ee
where $\delta_i=\delta_i(p)$ is a phase which could depend on the centre-of-mass momentum $p$ and $a_i, b_i$ are constants. Demanding that ${\cal S}_0$ and ${\cal S}_1$ project onto the singlet and the triplet channels, respectively, gives $a_1=3b_1$ and $a_0=-b_0$.  The overall normalization is fixed by the unitarity constraint, ${\cal S}^\dagger {\cal S}=1$, and we arrive at
\bea
\hspace{-.7cm}{\cal S} &=& \frac14 \left(3e^{i2\delta_1}+e^{i2\delta_0}\right) \bm{1} +\frac14  \left(e^{i2\delta_1}-e^{i2\delta_0}\right)\bm{\sigma}\cdot \bm{\sigma}  \, , \\
\hspace{-.7cm}&=&  \frac12\left(e^{2i\delta_1}+e^{2i\delta_0}\right) \bm{1} +\frac12\left(e^{2i\delta_1}-e^{2i\delta_0}\right) \text{\small SWAP} ,
\eea
where  we have used  $\text{\small SWAP} = (\bm{1}+ \bm{\sigma}\cdot \bm{\sigma})/2$. Surprisingly, rotational invariance and unitarity  constrains the $s$-wave $S$-matrix to be a linear combination of minimal entanglers. Realizing a minimally entangling gate in the $S$-matrix then requires the  relations: 
\begin{align}
 \delta_0(p) - \delta_1(p)&=0 &\Rightarrow & &{\cal S}&\sim [\bm{1}] \ , \\
| \delta_0(p)-\delta_1(p)|&=\pi/2 &\Rightarrow  & &{\cal S}&\sim \left[{\text{\small SWAP}}\right]  
\end{align}
Given the momentum dependence in $\delta_i$'s,  the above relations are not satisfied generically, except at certain fixed momenta. Therefore, when the relations hold independently of the momentum $p$, we are sitting in  special regions of parameter space. Indeed, these are  regions where symmetries emerge.

At sufficiently low energies, when the effective-range expansion is valid,  $p \cot\delta_i (p)$ is analytic in $p^2$
\be
p \cot\delta_i(p) = - \frac1{a_i} + \frac12 r_0^{(i)} p^2 +\cdots \ ,
\ee
where $a_i$ is the scattering length and $ r_0^{(i)}$ is the effective range. This translates into 
\be
\tan\delta_i(p) = - p a_i + {\cal O}(p^3) \ .
\ee
The limit $a_i \to 0$ and $\delta_i\to 0$ corresponds to a free theory in that channel, while $a_i\to \pm \infty$ and $\delta_i\to \mp\pi/2$ arises from the ``unitarity limit", with the largest possible cross section consistent with unitarity. Then it is clear that to achieve the $[\text{\small SWAP}]$ gate with $| \delta_0-\delta_1|=\pi/2$  for more than one $p$ one must tune $a\to \pm \infty$ in one channel and $a\to 0$ in the other channel. This corresponds to an RG fixed point of the theory, which is known to possess an emergent Schr\"odinger symmetry \cite{Mehen:1999nd}.

On the other hand, realizing the Identity with $\delta_0(p) =\delta_1(p)$ one needs to  set the scattering parameters in the two channels equal. This could be enforced through an internal symmetry, which we turn to next.

\section{Effective Hamiltonians} 

In this section we investigate what kind of Hamiltonians would give rise to the minimally entangling  $\left[\bm{1}\right]$ gate in the $S$-matrix. In a quantum field theory the phase $\delta_i$ is computed through the scattering amplitude ${\cal A}_i$ defined by $e^{2i\delta_i} = 1 + i (M p/2\pi){\cal A}_i$, where $M$ is the mass of the fermions (which we take to be degenerate, for simplicity). We then have
\be
{\cal A}_i = \frac{4\pi}{M} \frac{1}{p\cot \delta_i - i p} \ .
\ee
To realize the Identity gate we need $\delta_0=\delta_1$, which implies the amplitudes in both channels are equal, suggesting the existence of an internal symmetry. 

In what follows we will consider an effective Hamiltonians for two species of qubits,   labelled by the species index: $I,J=1,2$. Then the notation $|Ii;Jj\rangle$ represents the situation where Alice possesses a qubit $|i\rangle$ in species $I$  and Bob has a  qubit $|j\rangle$ in species $J$. The addition of the species quantum number opens the possibility of entanglement through the species quantum numbers.

An entangler that takes an input $|Ii;Jj\rangle$ to an output $|I^\prime i^\prime;J^\prime j^\prime\rangle$ can arise from the generic Hamiltonian,
\be
\bm{a}_{I^\prime i^\prime}^\dagger\,  \bm{a}_{Ii}\, \bm{a}_{J^\prime j^\prime}^\dagger\,  \bm{a}_{Jj}\ ,
\ee
where $\bm{a}^\dagger_{Ii}$  ($\bm{a}_{Ii}$) creates (annihilates) $|Ii\rangle$ and all other quantum numbers (such as the momentum) are suppressed.  The most general effective Hamiltonian is
\be
 H_{int} =\sum \ G_{ii^\prime jj^\prime} K_{II^\prime J J^\prime}\ \bm{a}_{I^\prime i^\prime}^\dagger\,  \bm{a}_{Ii}\, \bm{a}_{J^\prime j^\prime}^\dagger\,  \bm{a}_{Jj} + {\rm h.c.} \ ,
\ee
where $G_{ii^\prime jj^\prime}$ and $K_{II^\prime JJ^\prime}$ are coupling constants in the spin- and species-space, respectively.

From Eq.~(\ref{eq:idengate}) we see that, in order for $H_{int}$ to generate $\left[\bm{1}\right]$, it must be diagonal  in the spin-space, $G_{ii^\prime jj^\prime}= {\cal G}_{ij} \delta^{ii^\prime} \delta^{jj^\prime}$. Rotational invariance further implies ${\cal G}_{ij}$ must be independent of $(ij)$,
\be
\label{eq:rotG}
{\cal G}= g_s \left(\begin{array}{cc}
                            1 & 1 \\
                            1  &  1 \end{array}\right)\ ,
\ee
where $g_s$ is a constant. It turns out to achieve entanglement suppression when there is more than one species of qubit, $H_{int}$ must conserve the ``species number" for each individual species \cite{Beane:2018oxh}. To see this, consider the following species-changing one-body operators,
\be
\Theta = \bm{a}_{I\uparrow}^\dagger \bm{a}_{J\uparrow} + {\rm h.c.} \ .
\ee
When $\Theta$ acts on the state  $|I\uparrow;J\uparrow\rangle$, 
\be
\Theta|I\uparrow;J\uparrow\rangle = |I\uparrow;I\uparrow\rangle + |J\uparrow;J\uparrow\rangle\ ,
\ee
which will result in an entangled output. Therefore,  entanglement suppression  also requires $K_{II^\prime JJ^\prime} = {\cal K}_{IJ}\, \delta^{II^\prime}\delta^{JJ^\prime}$. If we use the notation $N_I=(\bm{a}_{I\uparrow}, \bm{a}_{I\downarrow})^T$, the Hamiltonian can now be written as 
\be
\label{eq:hint}
H_{int}= g_s  \sum_{I,J} {\cal K}_{IJ} \left(N_I^\dagger\,  N_I\right) \left(N_J^\dagger\,  N_J\right) + {\rm h.c.} \ ,
\ee
which is invariant under an $SU(2)_1\times SU(2)_2$ global symmetry,
\be
N_I \to U_I N_I\ , \qquad U_I \in U(2)_I\ , \  \  I=1,2\ .
\ee
The rotational invariance is the diagonal subgroup of  $SU(2)_1\times SU(2)_2$, which acts on each species identically.

In some cases it is possible to further assume the existence of a ``species symmetry" such that, much like the rotational invariance in Eq.~(\ref{eq:rotG}), the coupling constant ${\cal K}_{IJ}$ is universal and independent of $(IJ)$,
\be
{\cal K}= g_f \left(\begin{array}{cc}
                            1 & 1 \\
                            1  &  1 \end{array}\right)\ .
\ee
In the context of two-nucleon scattering the universality is achieved through the $SU(2)$ isospin symmetry transforming the neutron and the proton into each other. In this case  the Hamiltonian becomes
\be
\label{eq:genhint}
H_{int}= g_s g_f \left({\cal N}^\dagger\,  {\cal N}\right)^2\ ,
\ee
where ${\cal N}= (N_1, N_2)^T = (\bm{a}_{1\uparrow}, \bm{a}_{1\downarrow}, \bm{a}_{2\uparrow}, \bm{a}_{2\downarrow})^T$. We see the $SU(2)_1\times SU(2)_2$ global symmetry is enlarged to an $SU(4)$ symmetry. This is Wigner's $SU(4)$ spin-flavor symmetry in effective interactions of nucleons \cite{Mehen:1999qs}.

The above argument generalizes immediately to the case when there are $N_q$ species of qubit: $I, J=1, 2, \cdots, N_q$. In fact, Eq.~(\ref{eq:hint}) remains valid in the general case and the global symmetry is $N_q$-copy of $SU(2)$: $SU(2)_1\times \cdots SU(2)_{N_q}$. Moreover, when  there is a species-universal coupling constant (due to some underlying species symmetry), the Hamiltonian implementing the $[\bm{1}]$ gate  must be diagonal in both the spin- and species-space, with ${\cal N}$ in Eq.~(\ref{eq:genhint}) now defined by ${\cal N}= (\bm{a}_{1\uparrow}, \bm{a}_{1\downarrow}, \cdots, \bm{a}_{{N_q}\uparrow}, \bm{a}_{N_q\downarrow})^T$. The resulting global symmetry is   enlarged to $SU(2N_q)$. 

In low energy QCD interactions of the octet baryons indeed exhibit an enhanced $SU(16)$ spin-flavor symmetry \cite{Beane:2018oxh}. Now we see this is a direct consequence of QCD realizing the $[\bm{1}]$ gate.  The example also serves to demonstrate that, to arrive at a species-universal coupling constant among the $N_q$ qubit, it is not necessary to impose an $SU(N_q)$ species symmetry. In fact, it suffices if the $N_q$ species of qubit furnish an irreducible linear representation of an underlying group $G_q$ which may be smaller than $SU(N_q)$. For the octet baryon this is achieved through the $SU(3)$ flavor symmetry in QCD.

It is interesting to entertain the possibility of realizing the $[\text{\small SWAP}]$, which requires the $S$-matrix in one spin channel saturating the unitarity limit ($\delta=\pi/2$)  and the scattering being absent in the other channel ($\delta=0$).  Both $S$ matrices are invariant under the Schr\"odinger invariance~\cite{Mehen:1999nd}, which is not apparent at the level of the Lagrangian, but emerges after resumming certain contact interactions to all orders and performing a fine tuning of the theory parameters. For octet baryons the interactions in question are the two double trace operators in the chiral Lagrangian \cite{QiaofengThesis}. The resulting theory is not $SU(16)$ symmetric and  does not appear to be realized in low energy QCD.

\section{Conclusion}

In this work we initiated an information-theoretic study on the origin of symmetry. In particular, we studied the correlation between the emergence of global symmetries and entanglement suppression in the two-body scattering of fermions in the $s$-wave channel. After showing there are only two minimally entangling gates in a two-qubit system: the Identity and the ${\text{\small SWAP}}$ gates, which turned out to be rotationally invariant,  we argued the $S$-matrix  is constrained by unitarity and rotational invariance  to be a linear combination of the Identity and the ${\text{\small SWAP}}$.  Implementing a minimally entangling $S$-matrix requires certain conditions on the phases in the $S$-matrix in the $^1S_0$ and $^3S_1$ channels. The ${\text{\small SWAP}}$ gate can be realized when one of the channels flows to a free fixed point in the IR and the other a unitarity fixed point with the Schr\"odinger symmetry. On the other hand, an $S$-matrix implementing the Identity gate implies $[SU(2)]^{N_q}$ global symmetry in the effective Hamiltonian, where $N_q$ is the number of species of  qubits. When there is  a universal coupling among the species, the symmetry is enlarged to $SU(2N_q)$. 

In nature the Identity gate is realized in low energy QCD. One wonders whether the ${\text{\small SWAP}}$ gate arises naturally in other physical systems \footnote{An interesting possibility is the scattering of Kerr black holes, which also suppresses entanglement power \cite{Aoude:2020mlg}.}. Furthermore, while we focused on symmetries realized in the Wigner-Weyl mode \footnote{See Ref.~\cite{Cervera-Lierta:2017tdt} for a study on maximal entanglement and local symmetries.}, it would be interesting to study whether spontaneously broken symmetries, the Nambu-Goldstone mode, can  also be realized in an information-theoretic setting.

\noindent
{\bf Acknowledgement --}
I.~L. is supported in part by the U.S. Department of Energy under contracts No. DE- AC02-06CH11357 at Argonne and No. DE-SC0010143 at Northwestern. He also acknowledges the hospitalities at the National Center for Theoretical Sciences at National Taiwan University and National Tsing Hua University in Taiwan, where part of this work was performed. T.~M. acknowledges enlightening conversations with I. Marvian and Q. Liu. T.M. is supported by the U.S. Department of Energy, Office of Science, Office of Nuclear Physics under grant Contract Numbers  DE-FG02-05ER41368 and  DE-FG02-05ER41367.

\section{Appendix: Entanglement Measures}

This appendix gives a brief discussion of entanglement measures, which  we  argue can be motivated by demanding that they be invariant under unitary transformations on single qubits. The concurrence  \cite{PhysRevLett.80.2245}  turns out to be the simplest such invariant. We then discuss its relation to other  entanglement measures based on reduced density matrices, such as the von Neumann  entropy and the 2-entropy. Entanglement measures for operators are defined by applying them to 
unentangled direct product states and averaging over all such states. This is used to prove that all minimal entanglers are equivalent  to 
$\bm{1}$ or ${\text{\small SWAP}}$ up to local unitary operations.

For a two-qubit system the most general state is 
\bea
|\Psi\rangle =\alpha|00\rangle + \beta |01\rangle + \gamma|10\rangle + \delta|11\rangle \,.
\eea
Any entanglement measure should vanish on direct product states and attain a maximum on maximally entangled states. 
It is easy to show that $\Delta(\Psi)=2|\alpha\delta-\beta\gamma|$ defined  in Eq.~(\ref{eq:concur}) vanishes on all direct product states:
\bea
\label{eq:pure}
|\psi\rangle =(a|0\rangle+ b|1\rangle) \otimes (c|0\rangle+ d|1\rangle) \, .
\eea
In fact we will show that $\alpha \delta-\beta \gamma$ is the only second order polynomial in $\alpha, \beta,\gamma,$ and $\delta$ that has this property. 
It is also clear that $|\alpha \delta-\beta \gamma|  \leq |\alpha| |\delta|+  |\beta| |\gamma|  \leq (|\alpha|^2 +|\beta|^2+|\gamma|^2+|\delta|^2)/2 =1/2$,
so that $0 \leq \Delta(\Psi) \leq 1$. Finally $\Delta(\Psi)=1$ for Bell states, so $\Delta(\Psi)$ has the desired properties. 

Here we want to emphasize another way of motivating  $\Delta(\Psi)$, which could easily be generalized to $m$-level qubits. Under a unitary transformation $U_A$ on Alice the components 
of the wave function transform as:
\be
\left( \begin{array}{c} \alpha \\ \gamma \end{array}\right) \to U_A \left( \begin{array}{c} \alpha \\ \gamma \end{array}\right)\ ,\ \ 
\left( \begin{array}{c} \beta \\ \delta \end{array}\right) \to U_A \left( \begin{array}{c} \beta \\ \delta \end{array}\right) \ ,
\ee
while under a unitary transformation $U_B$ on Bob the components transform as 
\be
\left( \begin{array}{c} \alpha \\ \beta  \end{array}\right) \to U_B \left( \begin{array}{c} \alpha \\ \beta \end{array}\right) \ , \ \ 
\left( \begin{array}{c} \gamma \\ \delta \end{array}\right) \to U_B \left( \begin{array}{c} \gamma\\ \delta \end{array}\right) \ .
\ee
If we assemble the components of the wave function into a matrix
\bea
M\equiv \left( \begin{array}{cc} \alpha &\beta\\ \gamma & \delta \end{array}\right)\ ,
\eea
these transformations can be written as
\bea \label{trans}
M  \to U_A M U_B^T \, .
\eea
An important ingredient underlying this work is that local unitary transformations do not affect entanglement and this led us to the Cartan decomposition of a general $SU(4)$ element acting on two qubits. In the present context we see that entanglement measures must be invariants under the transformation in Eq.~(\ref{trans}). There are only two  invariants that are
second order: 
\bea 
{\rm Det}[M] &=& \alpha \delta -\beta \gamma \nn\\
{\rm Tr}[M^\dagger M] &=& |\alpha|^2 +|\beta|^2+|\gamma|^2+|\delta|^2 =1 \,.
\eea
Given this, it is not surprising that other entanglement measures for two qubits can be expressed in terms of $\Delta[\Psi]$. For example, the reduced density matrices for 
qubits $A$ and $B$ are:
\bea
\rho_A=M M^\dagger\ , \qquad \rho_B=M^\dagger M\,,
\eea
whose eigenvalues are
\bea
\lambda_\pm = \frac{1}{2}\left(1\pm \sqrt{1-\Delta^2(\Psi)}\right)\,.
\eea
Any entanglement measure derived from reduced density matrices can therefore be expressed in terms of the concurrence.
For example the von Neumann entropy is 
\bea
\!\!\!\!\!\! {\cal E}_E(|\Psi \rangle) = - {\rm Tr}[\rho_A \ln \rho_A] = -\lambda_+\ln \lambda_+ -  \lambda_-\ln \lambda_-\, ,
\eea
while
the 2-entropy  is
\bea 
\label{eq:ep2ent}
\!\!\!\!\!\! {\cal E}_R(|\Psi \rangle) =1-{\rm Tr}[ \rho_A^2] = \frac{1}{2}\Delta^2(\Psi)= 2 |\alpha \delta-\beta \gamma|^2\ .
\eea 
These share the property of $\Delta(\Psi)$ that they vanish when evaluated on an unentangled state. 
For maximally entangled states, ${\cal E}_E=\ln 2$ and ${\cal E}_R =1/2$. 

So far we have discussed entanglement measures for states. For operators the entanglement generated is dependent on the state they act on. For example, 
CNOT acting on the computational basis $\{|00\rangle ,|01\rangle, |10\rangle,|11\rangle\}$ yields  
$\{|00\rangle ,|01\rangle, |11\rangle,|10\rangle\}$ and generates no entanglement. But acting on the direct product state 
$(|0\rangle+|1\rangle)\otimes |0\rangle /\sqrt{2}$ the CNOT gate creates the maximally entangled state $|\Phi^+\rangle$.
Forming a sensible entanglement measure for operators requires some sort of averaging over initial states.
One possible approach is to average over all direct product states \cite{PhysRevA.63.040304,BallardWu2011}, which we will do below.

We parametrize the wave function of the qubits using Bloch sphere coordinates,
\bea 
a = \cos\frac{\theta_A}{2}\ , \quad b = \sin \frac{\theta_A}{2}e^{i\phi_A} \ ,\nn \\
c =  \cos\frac{\theta_B}{2}\ , \quad d = \sin \frac{\theta_B}{2}e^{i\phi_B}\ .
\eea
It is  simple to compute the concurrence for arbitrary $U_d$. Expanding the pure state $|\psi\rangle$ in Eq.~(\ref{eq:pure})
in terms of the Bell states, which are eigenstates of $U_d$, it is straightforward to show:
\bea \label{genconcurrence}
\left|\alpha \delta-\beta \gamma\right| &=& \frac{1}{4} \left| \sum_{i=1}^4 v_i \lambda_i^2 \right|  \ ,
\eea
where $\lambda_i=e^{i\phi_i}$ is defined through the $\phi_i$ in Eq.~(\ref{phases}) and 
\begin{align}
v_1 &= (ac+bd)^2 \ ,  & v_2 &=-(ac-bd)^2 \ ,\nn \\
v_3 &= -(ad+bc)^2 \ , &  v_4 &= (ad-bc)^2 \, .
\end{align}
For  averaging over direct product initial states it is more convenient to consider the entanglement power, which is the 2-entropy in Eq.~(\ref{eq:ep2ent}) averaged over all direct product states,    
\bea
\overline{{\cal E}_R(U_d|\psi\rangle)} &=& \overline{2 |\alpha \delta-\beta \gamma|^2} \nn \\
&=& \frac{1}{8} \overline{ \left| \sum_i v_i\lambda_i^2 \right|^2 }\nn \\
&=&  \frac{1}{8}  \sum_i \overline{ |v_i|^2}+ \frac{1}{8} \sum_{i\neq j} \overline{v_i v_j^*} \lambda_i^2 \lambda_j^{-2} \ ,
\eea 
where the last line follows from $\lambda_i^*=\lambda_i^{-1}$.  The overline denotes averaging over the solid angles $\Omega_{A,B}$. It is simple to evaluate the integral $\overline{v_i v_j^*}$,
\be
\overline{v_i v_j^*} = \int \frac{d\Omega_A}{4\pi} \int \frac{d\Omega_B}{4\pi} \ v_i v_j^* = -\frac{1}{9}+\frac{4}{9}\delta_{ij}\ ,
\ee
which yields~\cite{BallardWu2011,PhysRevA.70.052313}
\bea\label{epower}
&&\overline{ {\cal E}_R(U_d|\psi\rangle)} \nn \\
&&\  =  \frac{1}{6} -\frac{1}{18}\sum_{a<b} \cos(4\beta_a) \cos(4\beta_b)   \, .
\eea
The entanglement power is manifestly invariant  under the transformations $\beta_a\to \beta_a+\pi/2$, $(\beta_a,\beta_b)\to (\beta_b,\beta_a)$ and 
 $(\beta_a,\beta_b)\to (-\beta_a,-\beta_b)$ discussed in the main text. To have minimal entanglement we must have all $\cos(4\beta_a) \cos(4\beta_b)$ in Eq.~(\ref{epower}) 
 equal to $1$. This requires  $\beta_a= 0$ or $\pi/4$ for all $a$. The solution $\beta_a=0$ yields the identity gate,  $\bm{1}$,  and $\beta_a=\pi/4$ yields the  
 ${\text{\small SWAP}}$ gate times $e^{i \pi/4}$.  Thus we see that all minimal entanglers are either equivalent to $\bm{1}$ or ${\text{\small SWAP}}$ up to local unitary operations.

For $\beta_x=\beta_y=\beta_z\equiv\beta$, the entanglement power simplifies to
\be
\overline{ {\cal E}_R(U_d|\psi\rangle)} = \frac{\sin^2 4\beta}{6} \ .
\ee
In this case the eigenvalues of  $U_d$ are $e^{i\beta}$ and $e^{-3i\beta}$, while the eigenvalues of $S$ are $e^{2i\delta_1}$ and $e^{2i
\delta_0}$. Identifying $4 \beta$ with $2(\delta_1-\delta_0)$ we recover the entanglement power  derived in Ref.~\cite{Beane:2018oxh} .

\bibliography{QIS_ref}

\end{document}